\documentclass{aa}  

\usepackage{graphicx}
\usepackage{txfonts}

\newcommand{\gsim}{\raisebox{-0.6ex}{$\stackrel{{\displaystyle>}}{\sim}$}}

\begin{document}

\title{Search for $p$-mode oscillations in DA white dwarfs with VLT-ULTRACAM
\subtitle{I. Upper limits to the $p$-modes
\thanks{Based on observations obtained at the ESO Paranal Observatory
(programme 075.D-0371).}
}}

\author{R. Silvotti\inst{1}, G. Fontaine\inst{2}, M. Pavlov\inst{3}, 
T. R. Marsh\inst{4}, V. S. Dhillon\inst{5}, S. P. Littlefair\inst{5}, 
F. Getman\inst{6}
%
%
}

\offprints{R. Silvotti}

\institute{INAF--Osservatorio Astronomico di Torino, strada dell'Osservatorio
20, 10025 Pino Torinese, Italy\\
\email{silvotti@oato.inaf.it}
\and
D\'epartement de Physique, Universit\'e de Montr\'eal, C.P. 6128, 
Succ. Centre-Ville, Montr\'eal, Qu\'ebec H3C 3J7, Canada\\
\email{fontaine@astro.umontreal.ca}
\and
Sternberg Astronomical Institute, Universitatskij Prospect 13, Moscow, Russia\\
\email{pav@sai.msu.ru}
\and
Department of Physics, University of Warwick, Coventry CV4 7AL, UK\\
\email{t.r.marsh@warwick.ac.uk}
\and
Department of Physics and Astronomy, University of Sheffield, 
Sheffield S3 7RH, UK\\
\email{vik.dhillon@sheffield.ac.uk; s.littlefair@sheffield.ac.uk}
\and
INAF--Osservatorio Astronomico di Capodimonte, via Moiariello 16,
80131 Napoli, Italy\\
\email{tig@oacn.inaf.it}
}

\date{Received .... 2010 / Accepted .....}

\abstract
{}
{The main goal of this project is to search for $p$-mode oscillations 
in a selected sample of DA white dwarfs near the blue edge of the DAV ($g$-mode)
instability strip, where the $p$-modes should be excited following 
theoretical models.}
{A set of high quality time-series data on nine targets has been obtained 
in 3 photometric bands (Sloan u', g', r') using ULTRACAM at the VLT 
with a typical time resolution of a few tens of ms.
Such high resolution is required because theory predicts very short periods, 
of the order of a second, for the $p$-modes in white dwarfs.
The data have been analyzed using Fourier transform and correlation analysis 
methods.}
{$P$-modes have not been detected in any of our targets.
The upper limits obtained for the pulsation amplitude, typically less than
0.1\%, are the smallest limits reported in the literature.
The Nyquist frequencies are large enough to fully cover the frequency range 
of interest for the $p$-modes.
For the brightest target of our sample, G~185-32, a $p$-mode oscillation
with a relative amplitude of 5$\times 10^{-4}$ would have been easily 
detected, as shown by a simple simulation.
For G~185-32 we note an excess of power below $\sim$2 Hz in all the three 
nights of observation, which might be due in principle to tens of low-amplitude
close modes.
However, neither correlation analysis nor Fourier transform of the amplitude
spectrum show significant results.
We also checked the possibility that the $p$-modes have a very short lifetime, 
shorter than the observing runs, by dividing each run in several subsets
and analyzing these subsets independently.
The amplitude spectra show only a few peaks with S/N ratio higher than 
4~$\sigma$ but the same peaks are not detected in different subsets, as we 
would expect, and we do not see any indication of frequency spacing.

As a secondary result of this project, the detection of a new $g$-mode DAV 
pulsator near the blue edge of the ZZ~Ceti instability strip was claimed
(Silvotti et al. 2006) and will be described in detail in a forthcoming paper 
(Silvotti et al. 2010, paper II).
}
{}

\keywords{stars: white dwarfs -- stars: oscillations}

\authorrunning{R. Silvotti et al.}
\titlerunning{Search for $p$-mode oscillations in DA white dwarfs.
~I) Upper limits to the $p$-modes}

\maketitle


\section{Introduction}

From the first detection of a pulsating white dwarf (Landolt 1968), 
the number of known white dwarf pulsators has grown to the current 
number of almost two hundred, divided into three (plus perhaps one) distinct 
groups along the WD cooling sequence (see Fontaine \& Brassard 2008 and 
Winget \& Kepler 2008 for recent reviews).
All the pulsation periods detected so far in white dwarfs, with values between 
about 2 and 35 min, can be explained in terms of nonradial gravity mode
($g$-mode) oscillations.
These oscillations are well reproduced by adiabatic and nonadiabatic
theoretical models and white dwarf asteroseismology is capable of producing 
accurate measurements of mass, rotation period, thickness of the external 
layers of hydrogen or helium. 
An exhaustive review on white dwarf asteroseismology is given by Fontaine 
et al. (2010).

However, the same theory predicts that also acoustic modes ($p$-modes) could be
excited in white dwarfs.
These acoustic modes are mostly sensitive to the structure of the inner core 
(whereas the $g$-modes probe mainly the envelope) and have very short periods,
roughly between 0.1 and 10 seconds.
Radial ($p$-mode) oscillations in white dwarfs are predicted since a long time
(see Ostriker 1971 for a review of the early adiabatic work) and the first 
quasi-adiabatic or nonadiabatic theoretical studies have shown that some of
these modes should be excited (Vauclair 1971, Cox et al. 1980).
Using realistic atmospheric compositions for DA white dwarfs, computations
have shown that the blue edge of the radial instability strip lies at effective
temperatures slightly higher than for nonradial pulsations 
(Saio et al. 1983, Starrfield et al. 1983) and a similar situation occurs 
for DB white dwarfs (Kawaler 1993).
For both DA and DB white dwarfs, the maximum growth rates are obtained for high
overtone modes, with periods of a few tenths of second, with a lower limit 
near 0.1~s, set by the atmospheric acoustic cutoff (Saio et al. 1983, 
Hansen et al. 1985). For periods shorter than this limit, the reflective 
boundary condition at the surface is no longer valid.
Therefore the best period range to search for $p$-modes in white dwarfs should 
be between $\sim$0.1 and about 1~s, with longer periods up to about 12~s
(where the fundamental radial mode falls), that should have a lower 
observability due to their lower growth rates.
The best DA white dwarf candidates should be those near the DAV (or ZZ Ceti) 
$g$-mode instability strip and close to the radial blue edge, as the growth 
rates become smaller moving to the red.

From the observational side, high frequency $p$-mode pulsations have never 
been detected in white dwarfs.
Robinson (1984) reports the results on 19 DA white dwarfs (including 6 DAVs): 
five of them were observed with a photoelectric photometer at the 2.l~m 
McDonald telescope using a time resolution of 0.05 or 0.1 s; the other 14 
stars were observed in previous surveys with smaller telescopes.
The typical upper limits obtained were 1-2$\times$10$^{-3}$ (relative 
amplitude) with a Nyquist limit between 5 and 10~Hz.

Kawaler et al. (1994), with the High Speed Photometer aboard the HST, observed 
two DB white dwarfs with a time resolution of 10~ms: the DBV pulsator GD~358 
and PG~0112+104, a stable DB white dwarf with a slightly higher effective 
temperature.

Our new attempt is focused on a sample of nine DA white dwarfs near the blue 
edge of the DAV instability strip.
The combination of the large aperture of the VLT with the unique 
characteristics of ULTRACAM (high speed and high efficiency in 3 bands 
simultaneously) provides an ideal research tool for this project.

\section{Targets, observations and data reduction}

Most of the targets selected (six of them) are close to the blue edge of the 
DAV ($g$-mode) instability strip, according to theoretical expectations. 
Two of them, G~185-32 and L~19-2, are known DAV pulsators.
A third DAV pulsator, GD~133, has been discovered during our VLT run 
(Silvotti et al. 2006; 2010, paper II).
The other three targets lie in a larger range of effective temperatures, 
in order to check whether the $p$-mode instability strip might be shifted 
with respect to the expectations.

All the observations were performed in May 2005, when ULTRACAM was mounted 
for the first time at the visitor focus of the VLT UT3 ({\it Melipal}).
ULTRACAM\footnote{\it http://www.vikdhillon.staff.shef.ac.uk/ultracam/} 
(Dhillon et al. 2007) is a portable ultrafast 3-CCD camera that can reach 
a maximum speed of 300 frames per second in three photometric bands at the 
same time, selected among the {\it u'g'r'i'z'} filters of the Sloan Digital Sky 
Survey (SDSS) photometric system.

All the targets were observed using the SDSS filters $u'$, $g'$ and $r'$ during 
single runs with duration between 15 and 69 minutes.
Only the brightest target, G~185-32, was observed in three independent runs in
order to push down as much as possible the detection limit.
We used exposure times between 9 and 376~ms, depending on magnitude 
and sky conditions.
Further details concerning targets and observations are given in Table 1.

Near each target, at an angular distance between 1.1 and 1.9~arcmin,
we observed also a reference star in order to remove the spurious 
effects introduced by variable sky conditions.

Data reduction was carried out using the ULTRACAM pipeline (see Littlefair 
et al. 2008 for details).
After bias and flat field correction, we performed aperture photometry 
and we computed differential photometry dividing the target's counts
by the counts of the stable star.
Finally we applied the barycentric correction to the times.
In Fig.~1
the light curves of the nine targets are shown in a wide band
obtained summing the $g'$ and $r'$ counts.
This ``white light'' has a higher S/N ratio and for this reason it has been 
used in the first part of our analysis.

\begin{figure*}[t]
\label{lc}
\vspace{17.5cm}
\includegraphics{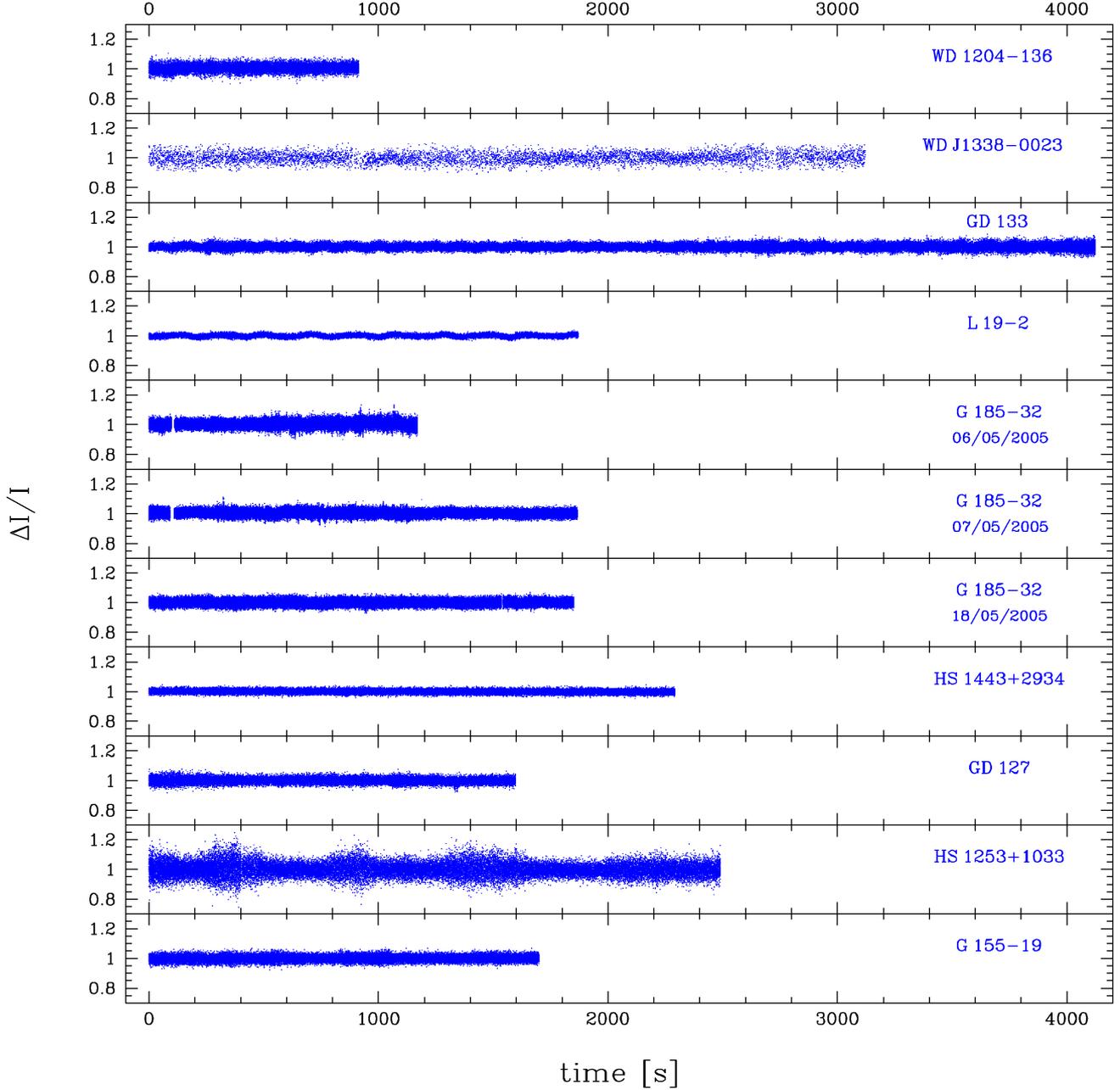}
\caption{Light curves of the eleven VLT runs in a wide photometric band 
obtained summing the $g'$ and $r'$ counts. The target's counts are divided
by the counts of the stable star.}
\end{figure*}

\subsection{Sky stability on short time scales using a 8m class telescope}

\begin{figure*}[th]
\label{scint}
\vspace{13.85cm}
\includegraphics{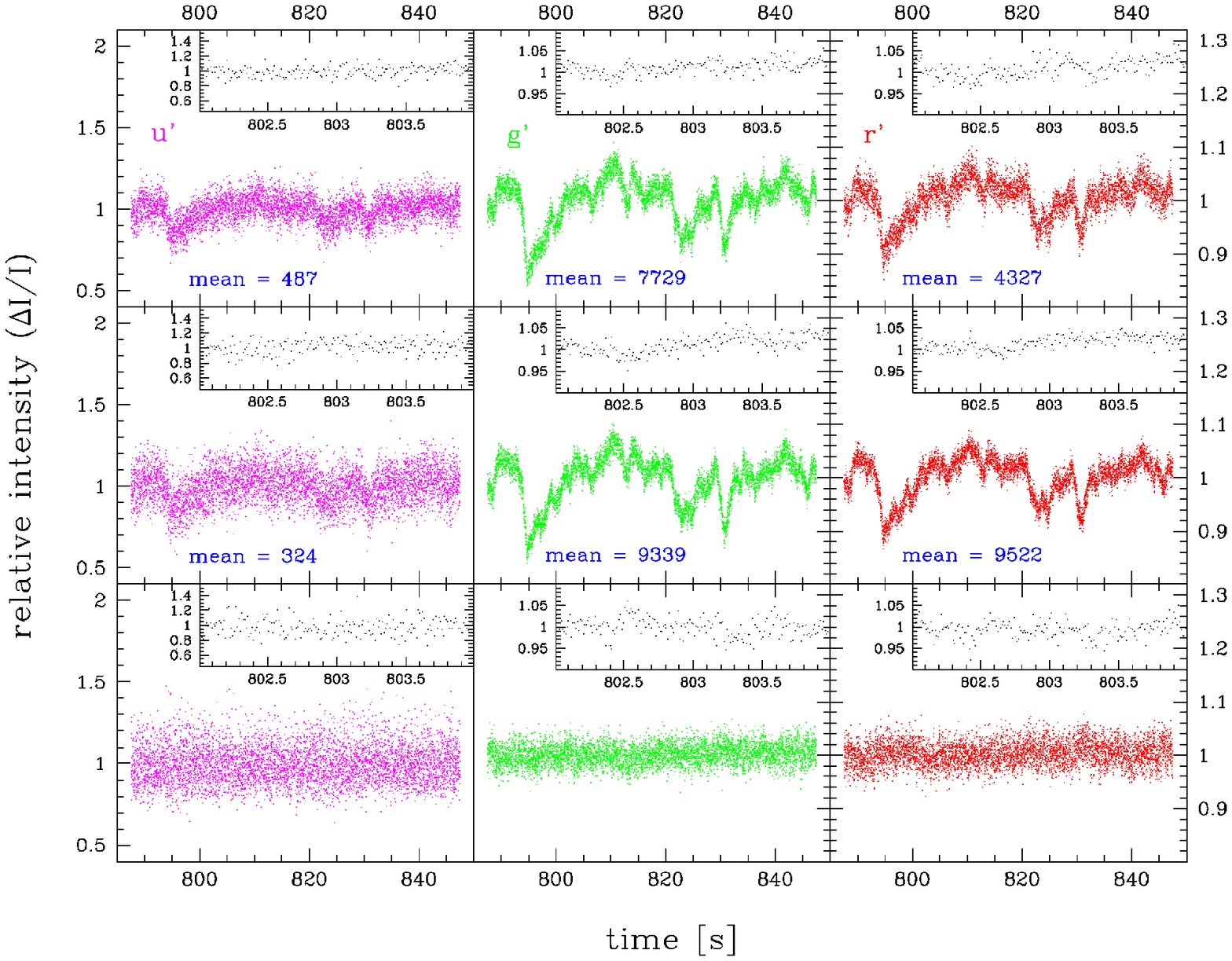}
\caption{Sky intensity fluctuations on time scales from tenths to tens of 
seconds ($u'$, $g'$ and $r'$ photometric bands).
A section of the light curve of G~185-32 (18 May 2005), with a duration  of 
1~min, is represented in the upper panels, while the reference star is shown 
in the central panels.
The average counts of each integration (8.8~ms) are reported.
The lower panels show the intensity ratio (target's counts divided by reference star's counts).
This figure shows that on time scales longer than about 1~s it is essential 
to have a reference star in order to reduce the sky fluctuations.
However, for shorter time scales, the light curves of the target and the
reference star are not coherent anymore, as we can see from the small panels
representing a short section (only 2~s) of the same light curves.
Note that the $u'$ band has a different vertical scale in all panels.}
\end{figure*}

In all our runs we see sky variations on time scales up to tens
of minutes that can reach amplitudes of 10\% and even more.
In our best runs on HS~1443+2934 and L~19-2 these variations have much
smaller intensities (few thousandths) and the flux remains constant within 
$\sim$3\% during the whole observation.
Our data allow to test the sky stability also on much shorter time scales, 
down to tenths of seconds.
An example is given in Fig.~2,
where we see that the variability
on time scales from seconds to tens of seconds resembles that of the typical
transparency variations on longer time scales (minutes to hours).
These variations are coherent at angular distances of 1-2~arcmin, as we
see comparing target and reference star's light curves.
However, at higher frequencies (\gsim~1~Hz), the spatial coherence is lost, as shown in the small panels of Fig.~2.

\section{Search for periodicities}

\begin{figure*}[t]
\label{as}
\vspace{17.5cm}
\includegraphics{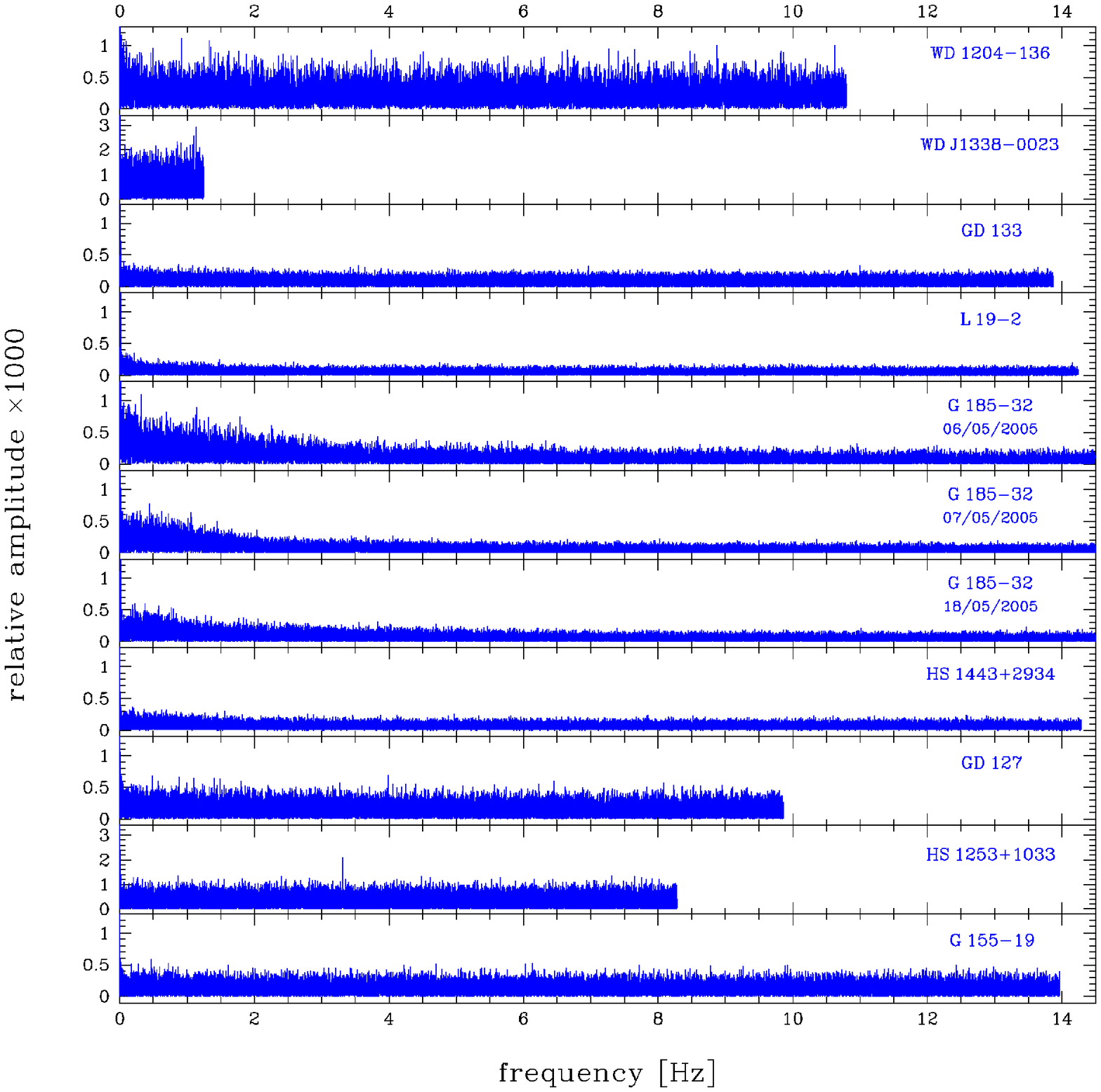}
\caption{Fourier amplitude spectra of the light curves of Fig.~1.
For G~185-32 the high frequency tail, from 14.5~Hz to the Nyquist frequency 
at about 46.7~Hz, is not shown in this plot and does not contain any 
significant peak.
The ordinate is in relative amplitude $\times$ 1000, which is equivalent to
milli-modulation amplitude units (mma, 1~mma=0.1\%=1000~ppm).
For two stars, WD~1338-0023 and HS~1253+1033, a different vertical scale 
has been used.
The peak near 3.3~Hz in the spectrum of HS~1253+1033\ 
is a false detection (see text for details).}
\end{figure*}

Fig.~3 shows the amplitude spectra of the eleven runs in the 
$g'$+$r'$ band. 
The only significant peak at about 3.3~Hz in the spectrum of HS~1253+1033\ 
is a false detection which appears also in the spectrum of the reference star 
alone.
These amplitude spectra were calculated using the flux ratios (target's counts 
divided by stable star's counts).
However, we have seen in Sect.~2.1 that differential photometry is useful only 
at frequencies lower than about 1 Hz (see Fig.~2 and Fig.~4).
For this reason we calculated also the amplitude spectra of the eleven runs
using only the target's counts. Again we did not detected any significant peak 
in any of the spectra. Negative results were found also when using the data of
the single $u'$, $g'$ and $r'$ photometric bands.

\begin{figure}[t]
\label{as_g185_32}
\vspace{9.5cm}
\includegraphics{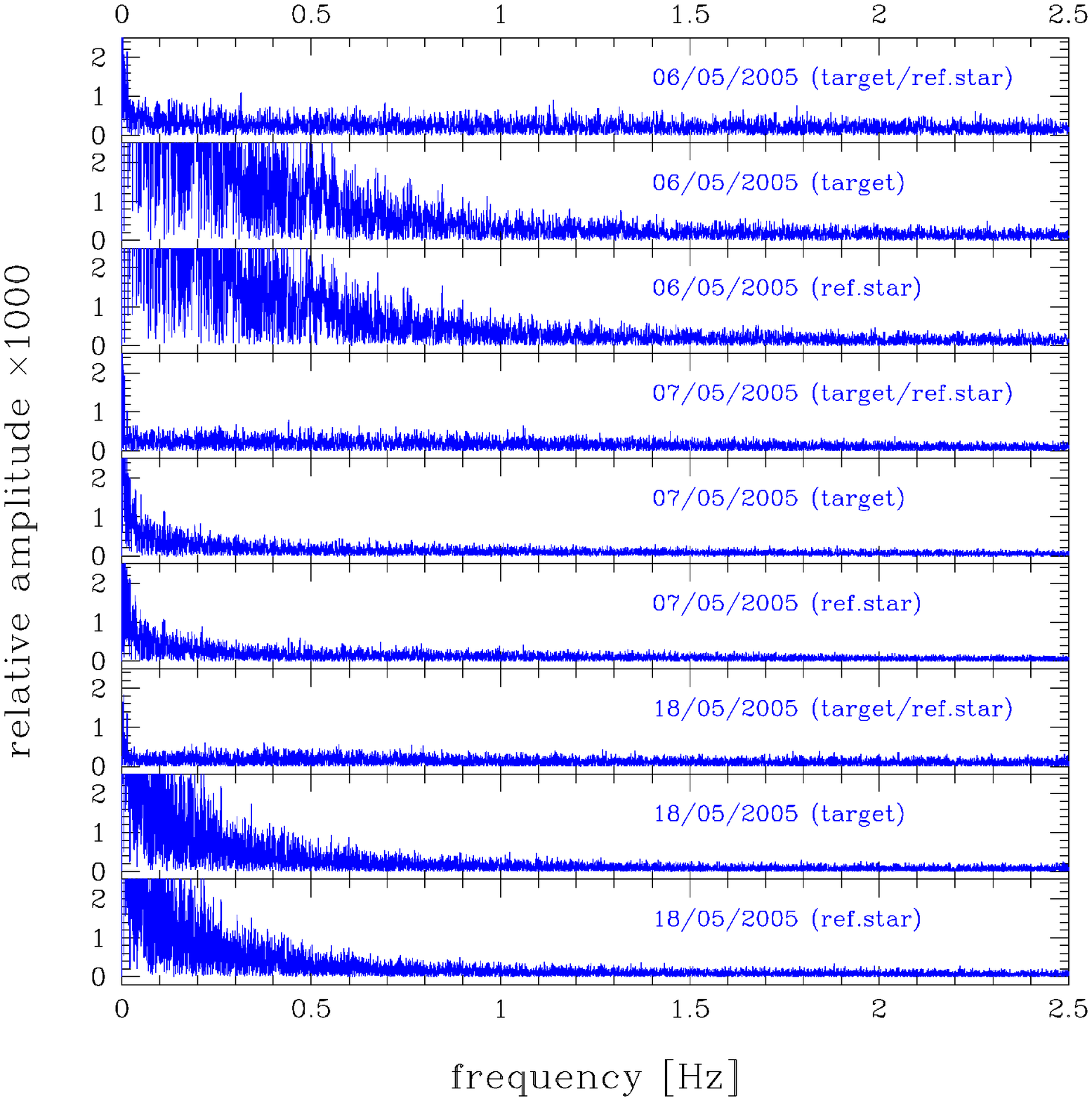}
\caption{Amplitude spectra of G~185-32 (detail of the low frequency part).
The amplitude spectra of the target and reference star alone are also reported.
As already seen in Fig.~2,
these plots confirm that above about 1~Hz 
(where transparency variations cease) the mean noise of the amplitude spectrum 
is lower when we consider the target alone instead of using the flux 
ratio.}
\end{figure}

Looking at Fig.~3,
we note an excess of power below $\sim$2 Hz
in all the three nights of observation on G~185-32, the brightest target 
of our sample.
This excess, which is only marginally present in three other targets
(L~19-2, GD~133 and HS~1443+2934), could be due in principle to tens of close 
modes with low amplitude.
However, when we look in detail to the low frequency part of the spectra, 
we see that the low-amplitude peaks are different from night to night, 
contrary to what we would expect (Fig.~4).
To test this hypothesis in greater depth, we can use one of the properties 
of the $p$-modes: the high-overtone modes should be almost equally spaced in 
frequency (a constant spacing is expected when the adiabatic sound speed is 
constant in a star with an homogeneous composition).
An almost constant frequency spacing would be easily detected using correlation
analysis or through Fourier transform (FT) of the amplitude spectrum 
(or FT$^2$).
For all our targets we have used both methods without detecting any significant
peak. An example is given in Fig.~5,
where FT$^2$ and 
autocorrelation function of G~185-32 are shown and compared with those of a 
synthetic run containing a set of 20 equally spaced frequencies between 
1.265 and 2.5~Hz with same amplitude.
(see caption of Fig.~5
for more details).
With relative amplitudes equal to 2$\times 10^{-4}$, no synthetic sinusoids 
are detected in the amplitude spectrum. With larger amplitudes of 
4$\times 10^{-4}$ (6$\times 10^{-4}$), 8 out of 20 (20/20) sinusoids
are found.
With 4$\times 10^{-4}$ (6$\times 10^{-4}$) a significant peak is found also 
from correlation analysis at 6~$\sigma$ (17~$\sigma$), while FT$^2$ 
is less sensitive.

\begin{figure*}[t]
\label{ft2-acf}
\vspace{13.5cm}
\includegraphics{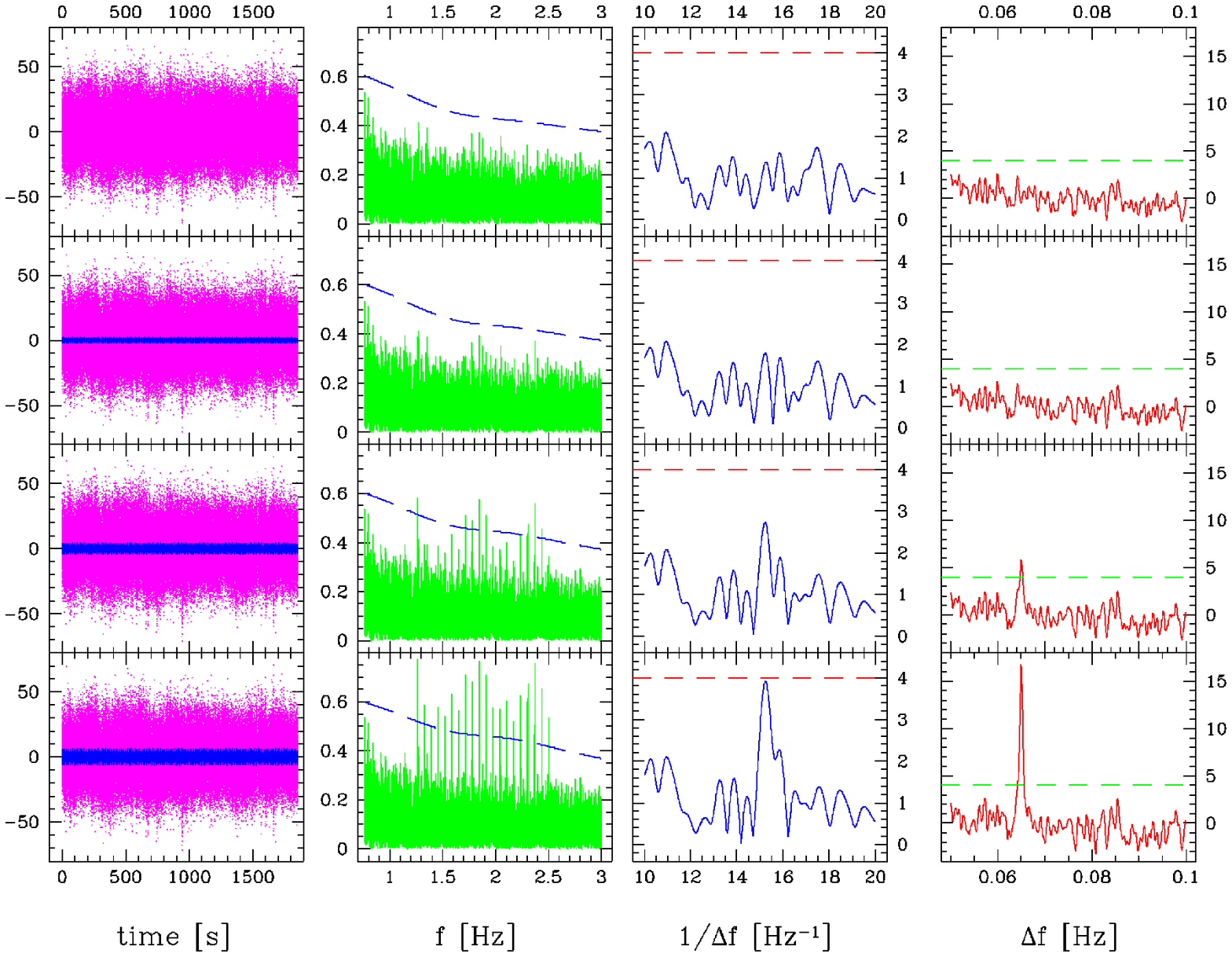}
\caption{Comparison between G~185-32 (18 May 2005, upper panels) and synthetic 
data.
Each synthetic dataset was constructed by adding to the VLT data a sequence 
of 20 sinusoidal waves between 1.265 and 2.5~Hz, equally spaced in frequency 
($\Delta$f=0.065~Hz).
All the 20 sinusoids have same relative amplitude of 2$\times 10^{-4}$, 
4$\times 10^{-4}$ and 6$\times 10^{-4}$ respectively.
The various columns represent respectively: light curve with synthetic sinusoids
superimposed (1), amplitude spectrum (2), amplitude spectrum of the amplitude 
spectrum or FT$^2$ (3), autocorrelation function of the amplitude spectrum (4).
The ordinate is given in relative amplitude $\times$ 1000 for column 1 and 2, 
and in $\sigma$ units for column 3 and 4.
The dashed lines show the 4$\sigma$ detection threshold, which correspond
to 4 times the average noise for column 2 and 3 and to 4 times the standard
deviation for column 4.
In column 2 the mean noise was calculated using a (smooth) cubic spline 
interpolation in order to consider also the noise increase at low frequency 
due to transparency variations.}
\end{figure*}

Another possibility that we have considered is that the $p$-modes are not
seen because of short life times.
With life times shorter than the observing runs, their amplitude in the Fourier
spectrum would be reduced due to phase incoherence and they could escape 
detection.
In order to verify also this possibility, we have divided each run in a number
of subruns of at least 14~s each and calculated the amplitude spectrum of each 
subset. 
Then, an automated procedure has found all the peaks with an amplitude higher 
than 
3 times the local noise.
The local noise has been determined by fitting each amplitude spectrum with a 
(smooth) cubic spline; an example of cubic spline interpolation is shown
in Fig.~5 (2$^{nd}$ column).
This procedure has been repeated for different size of the subsets, starting 
from 1,344 data points and increasing the size by a factor of $\sim$2 at each
iteration.
The results for the three runs on G~185-32 are illustrated in Fig.~6.
We see from this plot that only a few peaks with S/N ratio higher than 
4~$\sigma$ are found but, also in this case, the same peaks are not detected 
in different subsets, as we would expect.
Moreover, from autocorrelation analysis we do not obtain any indication of
frequency spacing in these subsets.
We conclude that also the possibility that the $p$-modes have very short life 
times is not supported by our observations.
This conclusion is valid for all the targets of our sample.

\begin{figure}[t]
\label{sub}
\vspace{9.5cm}
\includegraphics{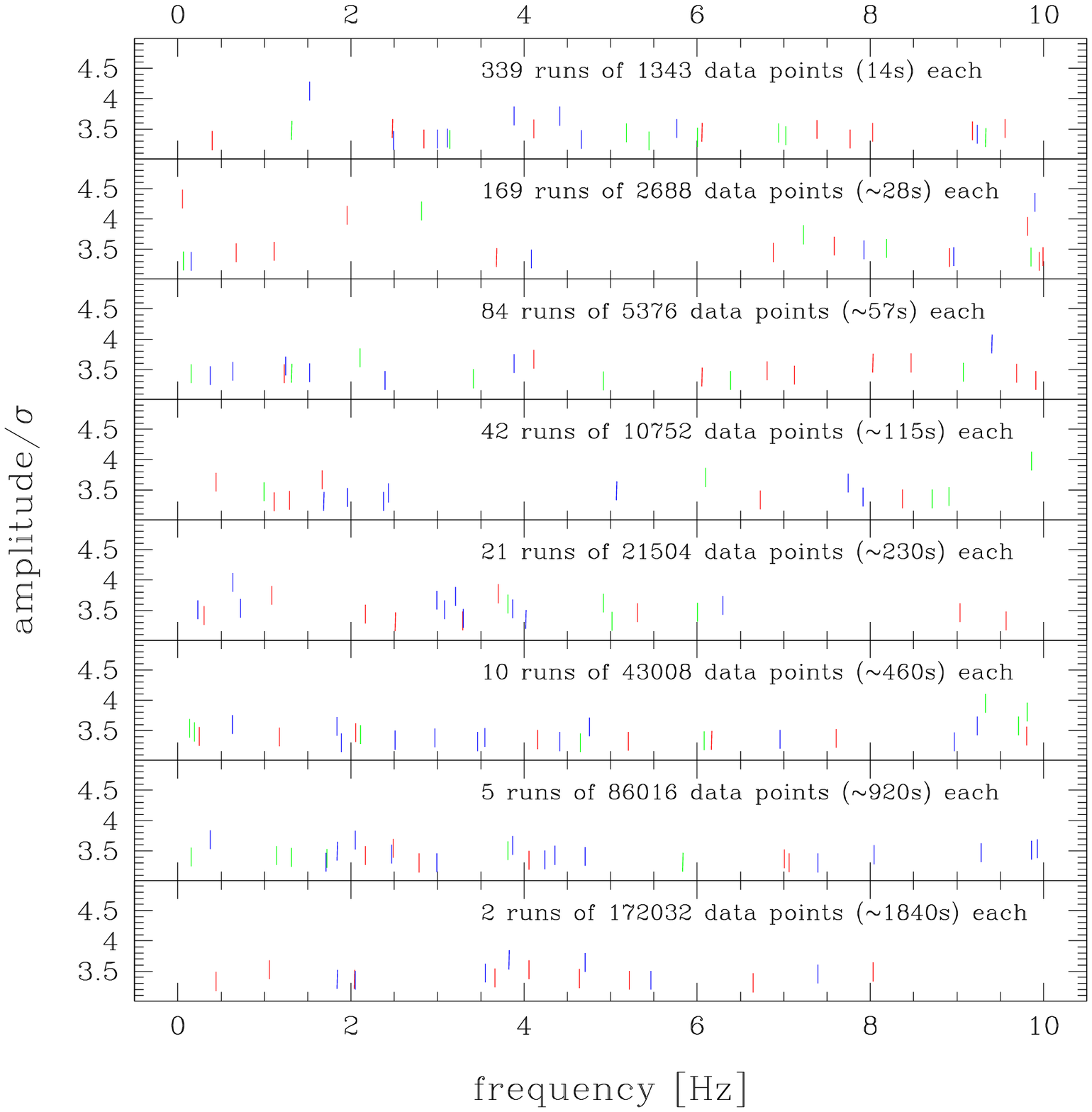}
\caption{Search for pulsations with life times shorter than the observing runs:
each panel shows the peaks with amplitude higher than 3.3$\sigma$, detected 
in $n$ subsets with same length, extracted from the three runs on G~185-32.
Note that the detection limit varies from panel to panel as it goes with 
$d^{-1/2}$, where $d$ is the duration of each subset.
In this plot, the 4$\sigma$ detection threshold goes from a relative amplitude 
of $\sim$$5 \times 10^{-4}$ (lower panel) to about $6 \times 10^{-3}$ 
(upper panel). Only for the online electronic version, the different 
colors refer to the different dates of the runs: green=6 May, red=7 May, 
blue=18 May 2005.}
\end{figure}

\begin{table*}[t,h]
\caption{Targets, observations and upper limits to the $p$-modes}
\smallskip
\begin{center}
{\small
\begin{tabular}{llcccccccccc}
\hline
\noalign{\smallskip}
\noalign{\smallskip}
Name(s) & & \hspace{-2.7mm} B   & \hspace{-3.2mm} T$_{\it eff}$ & \hspace{-3.5mm} log$g$ & \hspace{-2mm} Ref & T$_{0}^{1}$ & \hspace{-1mm} length & $\Delta$t$^{2}$ & \hspace{-2mm} Nyquist & \hspace{-2mm} A1$_{lim}^{3}$     & \hspace{-3mm} A2$_{lim}^{4}$\\
        & & \hspace{-2.5mm} mag & \hspace{-3.2mm} [K]           & \hspace{-3.5mm} [cgs]  &                   & [days]      & \hspace{-1mm} [s]    & [ms]            & \hspace{-2mm} [Hz]    & \hspace{-2mm} ($\times$10$^{4}$) & \hspace{-3mm} ($\times$10$^{4}$)\\
\noalign{\smallskip}
\hline
\noalign{\smallskip}
\noalign{\smallskip}
WD~1204--136    & \hspace{-3mm} EC~12043--1337 & \hspace{-3mm} 15.7 & \hspace{-1.7mm} 11,180$^a$ & \hspace{-2mm} 8.24$^a$   & \hspace{-2mm} 1 & 20.00474258 & \hspace{-0.mm} 914 & 46 &  \hspace{-2.6mm} 10.87 & \hspace{-2mm} 10.9 & \hspace{-3mm} 10.1\\
WD~J1338--0023  & \hspace{-3mm} WD 1335--001   & \hspace{-3mm} 17.3 & \hspace{-1.7mm} 11,980$^b$ & \hspace{-2.4mm} 7.94$^b$ & \hspace{-2mm} 1 & 20.95120770 & \hspace{-1.5mm} 3121 & \hspace{-2.4mm} 441 & 1.13 & \hspace{-2mm} 29.2 & --\\
WD~1116+026$^5$ & \hspace{-3mm} GD~133         & \hspace{-3mm} 14.8 & \hspace{-3mm} 12,090 & \hspace{-3.8mm} 8.06           & \hspace{-2mm} 1 & 15.95335902 & \hspace{-1.5mm} 4124 &  37 & \hspace{-2.6mm} 13.51 &  \hspace{-0.4mm} 3.6 &  \hspace{-1.4mm} 3.4 \\
WD~1425--811$^5$& \hspace{-3mm} L~19-2         & \hspace{-3mm} 14.0 & \hspace{-3mm} 12,100 & \hspace{-3.8mm} 8.21           & \hspace{-2mm} 2 & 11.25617640 & \hspace{-1.5mm} 1870 &  35 & \hspace{-2.6mm} 14.29 &  \hspace{-0.4mm} 3.5 &  \hspace{-1.4mm} 2.1 \\
WD~1935+276$^5$ & \hspace{-3mm} G~185-32       & \hspace{-3mm} 13.1 & \hspace{-3mm} 12,130 & \hspace{-3.8mm} 8.05           & \hspace{-2mm} 2 & 06.42004750 & \hspace{-1.5mm} 1170 &  11 & \hspace{-2.6mm} 45.45 & \hspace{-2mm} 10.9 &  \hspace{-1.4mm} 6.0 \\
                &                              &                    & 			     &		                    &   & 07.40988878 & \hspace{-1.5mm} 1867 &  11 & \hspace{-2.6mm} 45.45 &  \hspace{-0.4mm} 7.8 &  \hspace{-1.4mm} 2.7 \\
                &                              &                    & 			     &		                    &   & 18.28107923 & \hspace{-1.5mm} 1850 &  11 & \hspace{-2.6mm} 45.45 &  \hspace{-0.4mm} 5.9 &  \hspace{-1.4mm} 3.6 \\
HS~1443+2934    &                              & \hspace{-3mm} 14.5 & \hspace{-3mm} 12,400 & \hspace{-5.0mm} 8.1            & \hspace{-2mm} 3 & 11.22092258 & \hspace{-1.5mm} 2293 &  35 & \hspace{-2.6mm} 14.29 &  \hspace{-0.4mm} 3.6 &  \hspace{-1.4mm} 2.5 \\
WD~J1105+0016   & \hspace{-3mm} GD~127         & \hspace{-3mm} 15.4 & \hspace{-3mm} 12,850 & \hspace{-3.6mm} 8.26           & \hspace{-2mm} 4 & 17.01330012 & \hspace{-1.5mm} 1598 &  51 &  9.80 &  \hspace{-0.4mm} 6.8 &  \hspace{-1.4mm} 6.9 \\
HS~1253+1033    &                              & \hspace{-3mm} 14.4 & \hspace{-1.7mm} 13,040$^c$ & \hspace{-2.4mm} 7.85$^c$ & \hspace{-2mm} 5 & 16.19138485 & \hspace{-1.5mm} 2490 &  57 &  8.77 & \hspace{-2mm} 13.4 & \hspace{-1.4mm} 13.4$^6$ \\
WD~1827-106     & \hspace{-3mm} G~155-19       & \hspace{-3mm} 14.4 & \hspace{-3mm} 13,300 & \hspace{-3.8mm} 7.63           & \hspace{-2mm} 1 & 16.22905171 & \hspace{-1.5mm} 1701 &  36 & \hspace{-2.6mm} 13.89 &  \hspace{-0.4mm} 5.8 &  \hspace{-1.4mm} 5.2 \\
\noalign{\smallskip}
\hline
\end{tabular}}
\end{center}
\label{obslog}
$^1$ JD of the 1$^{st}$ datum -- 2453491.5 (2453491.5 = 1$^{st}$ of May 2005, 0~h UT).\\
$^2$ Time resolution.\\
$^3$ Relative amplitude of the highest peak in the Fourier spectrum for ~0.1~Hz$<$f$<$2~Hz.\\
$^4$ Relative amplitude of the highest peak in the Fourier spectrum for ~2~Hz$<$f$<$Nyquist.\\
$^5$ ZZ Ceti pulsator.\\
$^6$ Excluding the spurious peak at $\sim$3.3~Hz (see section 3).\\
$^a$ T$_{\it eff}$=11,110 and log$g$=8.05 from Koester et al. 2001.\\
$^b$ T$_{\it eff}$=11,650 and log$g$=8.08 from Mukadam et al. 2004.\\
$^c$ T$_{\it eff}$=12,600 and log$g$=8.5 from Silvotti et al. 2005.\\
Ref: 1: Gianninas et al. 2005; 2: Bergeron et al. 2004; 3: Silvotti et al. 2005; 4: Mukadam et al. 2004; 5: Gianninas et al. 2007.

%
%
%
%

\end{table*}

\section{Summary}

We have not detected any significant peak in 
the amplitude spectra of nine DA white dwarfs near the DAV instability strip
in the range of frequencies expected for the $p$-modes.
The upper limits that we have obtained for the pulsation amplitudes are 
reported in Table 1, together with the main characteristics of the stars 
observed.
Thanks to the high efficiency of the VLT-ULTRACAM system, our results move 
down the detection limit for the $p$-modes in DA white dwarfs to less than 
0.1\% for most of our targets.
These limits are lower by a factor of about 2-3 (at same magnitude level) 
respect to previous findings (Robinson 1984) and the Nyquist frequencies are 
much larger, covering the whole range of frequencies expected for the $p$-modes.
For G~185-32, the brightest target of our sample, an apparent excess of power 
is seen below $\sim$2~Hz in all the three nights of observation but none 
of our analysis allows us to deduce that this apparent excess of power is 
indicative of the presence of $p$-modes.
As shown by a simple simulation, a peak with a relative amplitude of 
6$\times 10^{-4}$ (3$\times 10^{-4}$) in the frequency range 1-3~Hz
(3-10~Hz) would have been easily detected in the amplitude spectrum of
G~185-32.

In our analysis we have considered various possibilities that could hide
the $p$-modes (low amplitudes, very short life times) and applied various 
techniques that can help to bring the signal out of the noise.
Using one of the properties of the high overtones $p$-modes, that should be 
almost equally spaced in frequency, we have searched for a constant 
frequency spacing through correlation analysis and Fourier transform of the 
amplitude spectrum (FT$^2$), without finding significant results.
Then we have divided each data set into several subsets of varying length
and analyzed each subset independently, again without finding any trace of
excited modes with very short life times.

Our results do not necessarily mean that the $p$-modes in DA white dwarfs stars
are not excited at all.
As the $p$-modes act mainly in the vertical direction, and vertical motions are
limited by the huge gravity, a very low amplitude, below our detection limit,
would not be surprising.
We will examine in a subsequent publication with the help of detailed models 
how this can come about.
The upper limits reported in this paper can help to constrain the nonadiabatic 
models of the DA white dwarfs, which have large uncertainties because of 
the convection.


\begin{acknowledgements}

ULTRACAM is supported by STFC grant PP/D002370/1. 
S.P.L. acknowledges the support of an RCUK Fellowship and STFC grant 
PP/E001777/1.
G.F. acknowledges the contribution of the Canada Research Chair Program.
A first observational search for $p$-mode pulsations in DA white dwarfs started
in 2003, based on two observing runs at the SAO 6~m telescope using the 
MANIA instrument.
Because of bad weather conditions, both the observing runs did not produced 
useful data.
R.S. and M.P. wish to thank Grigory M. Beskin, Sergey V. Karpov and Vladimir L.
Plokhotnichenko for the collaboration to the observations and for their
kind hospitality at the 6~m SAO telescope in June 2004.
Finally we thank the referee, Susan E. Thompson, for a careful reading of the 
manuscript and for useful suggestions.





\end{acknowledgements}



\begin{thebibliography}{}

\bibitem[2004]{bergeron04}
Bergeron, P., Fontaine, G., Bill\`eres, M., Boudreault, S., Green, E.~M., 2004,
\apj~ 600, 404

\bibitem[1980]{cox80}
Cox, A.~N., Hodson, S.~W., Starrfield, S.~G., 1980, Proc. of the workshop on 
``Nonradial and nonlinear stellar pulsation'', Springer-Verlag, Berlin, p.458

\bibitem[2007]{dhillon07}
Dhillon, V.~S., Marsh, T.~R., Stevenson, M.~J., et al., 2007, MNRAS 378, 825




\bibitem[2008]{fontaine_brassard_08}
Fontaine, G., Brassard, P., 2008, PASP 120, 1043

\bibitem[2010]{fontaine10}
Fontaine, G., Brassard, P., Charpinet, S., Quirion, P.-O., Randall, S.~K., 2010,
in ``White Dwarfs'', Monograph Series sponsored by the Royal Astronomical Society, ed. R. Napiwotzki \& M. Burleigh, Springer, in press

\bibitem[2005]{gianninas05}
Gianninas, A., Bergeron, P., Fontaine, G., 2005, ApJ 631, 1100

\bibitem[2007]{gianninas07}
Gianninas, A., Bergeron, P., Fontaine, G., 2007, ASP Conf. Series 372, 577

\bibitem[1985]{hansen85}
Hansen, C.~J., Winget, D.~E., Kawaler, S.~D., 1985, ApJ 297, 544

\bibitem[1993]{kawaler93}
Kawaler, S.~D., 1993, ApJ 404, 294

\bibitem[1994]{kawaler94}
Kawaler, S.~D., Bond, H.~E., Sherbert, L.~E., Watson, T.~K. 1994, AJ 107, 298



\bibitem[2001]{koester01}
Koester, D., Napiwotzki, R., Christlieb, N., et al., 2001, A\&A 378, 556

\bibitem[1968]{landolt68}
Landolt, A.~U., 1968, ApJ 153, 151

\bibitem[2008]{littlefair08}
Littlefair, S.~P., Dhillon, V.~S., Marsh, T.~R., et al., 2008, MNRAS 391, L88

\bibitem[2004]{mukadam04}
Mukadam, A.~S., Mullally, F., Nather, R.~E., et al., 2004, ApJ 607, 982

\bibitem[1971]{ostriker71}
Ostriker, J.~P., 1971, Ann. Rev. A\&A 9, 353

\bibitem[1984]{robinson84}
Robinson, E.~L. 1984, AJ 89, 1732

\bibitem[1984]{saio83}
Saio, H., Winget, D.~E., Robinson, E.~L., 1983, ApJ 265, 982

\bibitem[2005]{silvotti05}
Silvotti, R., Voss, B., Bruni, I., et al. 2005 A\&A 443, 195

\bibitem[2006]{silvotti06}
Silvotti, R., Pavlov, M., Fontaine, G., Marsh, T., Dhillon, V., 2006, MmSAI 77, 486

\bibitem[2010]{silvotti10}
Silvotti, R., Pavlov, M., Fontaine, G., Marsh, T.~R., Dhillon, V.~S., 
Littlefair, S.~P., 2010, A\&A in preparation (paper II)

\bibitem[1983]{starrfield83}
Starrfield, S., Cox, A.~N., Hodson, S.~W., Clancy, S.~P., 1983, ApJ 269, 645 

\bibitem[1971]{vauclair71}
Vauclair, G., 1971, Proc. of the IAU Symposium 42 on ``White Dwarfs'', edited 
by W.J. Luyten, Springer-Verlag, Dordrecht, p.145

\bibitem[2008]{winget_kepler_08}
Winget, D.~E., Kepler, S.~O., 2008, Ann. Rev. A\&A 46, 157

\end{thebibliography}
\end{document}